\documentclass[11pt]{article}
\usepackage{amssymb}
\usepackage{amsfonts}
\usepackage{amsmath}

\def\a{\alpha}
\def\b{\beta}
\def\g{\gamma}

\def\eps{\epsilon}

\def\z{\zeta}

\def\t{\theta}

\def\n{\nu}

\def\s{\sigma}
\def\p{\phi}

\def\c{\raise2pt\hbox{$\chi$}}

\def\dt#1{\dot{#1}}
\def\ddt#1{\ddot{#1}}
\def\ad{{\dt{\alpha}}}
\def\bd{{\dt{\beta}}}
\def\gd{{\dt{\gamma}}}
\def\add{{\ddt{\alpha}}}
\def\bdd{{\ddt{\beta}}}

\def\zd{{\dt{2}}}
\def\od{{\dt{1}}}
\def\zdd{{\ddt{2}}}
\def\odd{{\ddt{1}}}

\def\Db{{\bar{D}}}
\def\ct{{\tilde{\c}}}
\def\zb{{\bar{\zeta}}}
\def\zh{{\hat{\zeta}}}

\def\sfrac#1#2{{\textstyle\frac{#1}{#2}}}

\def\+{\dagger}
\def\={\ =\ }
\def\pa{\partial}
\def\pab{{\bar{\partial}}}
\def\>{\rangle}
\def\<{\langle}

\def\nzero{${\cal N}{=}0$}

\def\nfour{${\cal N}{=}4$}

\newcommand{\im}{\,\mathrm{i}\,}
\newcommand{\diff}{\mathrm{d}}
\newcommand{\R}{{\mathbb{R}}}
\newcommand{\C}{{\mathbb{C}}}

\newcommand{\tr}{{\mathrm{tr}}}

\newcommand{\cP}{{\mathcal P}}
\newcommand{\cA}{{\mathcal A}}
\newcommand{\cL}{{\mathcal L}}
\newcommand{\cO}{{\mathcal O}}
\newcommand{\cT}{{\mathcal T}}

\newcommand{\beq}{\begin{equation}}
\newcommand{\eeq}{\end{equation}}
\newcommand{\bea}{\begin{eqnarray}}
\newcommand{\eea}{\end{eqnarray}}
\newcommand{\bal}{\begin{aligned}}
\newcommand{\eal}{\end{aligned}}

\newcommand{\und}{\qquad\text{and}\qquad}

\begin{document}
\begin{titlepage}
\setcounter{page}{0}
\begin{flushright}
hep-th/0406179\\
ITP--UH--19/04
\end{flushright}

\vspace{10mm}

\begin{center}

{\LARGE\bf Supertwistors and Cubic String Field Theory \\[8pt]
           for Open N=2 Strings}

\vspace{20mm}

{\Large Olaf Lechtenfeld \ \ and \ \ Alexander D. Popov~$^*$} \\[12pt]
{\em Institut f\"ur Theoretische Physik, Universit\"at Hannover \\
     Appelstra\ss{}e 2, D-30167 Hannover, Germany}\\[6pt]
{\tt lechtenf, popov @itp.uni-hannover.de}

\vspace{30mm}

\begin{abstract}
\noindent 
The known Lorentz invariant string field theory for open $N{=}2$ strings
is combined with a generalization of the twistor description of 
anti-self-dual (super) Yang-Mills theories. We introduce a Chern-Simons-type
Lagrangian containing twistor variables and derive the Berkovits-Siegel
covariant string field equations of motion via the twistor correspondence.
Both the purely bosonic and the maximally space-time supersymmetric cases 
are considered.
\end{abstract}

\end{center}

\vfill

\textwidth 6.5truein
\hrule width 5.cm
\vskip.1in

{\small
\noindent ${}^*$
On leave from Laboratory of Theoretical Physics, JINR,
Dubna, Russia}

\end{titlepage}

\section{Introduction}

It was recently shown by Witten~\cite{Witten} that B-type open topological 
string theory with the supertwistor space $\C P^{3|4}$ as a target space is 
equivalent to holomorphic Chern-Simons (hCS) theory on the same space 
(for related works see~\cite{Berkovits}--\cite{BeWi}). 
This hCS theory in turn is equivalent to supersymmetric
\nfour\ anti-self-dual Yang-Mills (ASDYM) theory in four dimensions. The 
\nfour\ super ASDYM model is governed by the Siegel Lagrangian~\cite{Si92}. 
Its truncation to the bosonic sector describes \nzero\ ASDYM theory with an 
auxiliary field of helicity $-1$~\cite{Si92,ChaSi96}.

It may be of interest to generalize the twistor correspondence to the level 
of string field theory (SFT). This could be done using the approach proposed
in~\cite{Motl} or in the more general setting of~\cite{Siegel}. 
Alternatively, one could concentrate on (an appropriate extension of) SFT 
for $N{=}2$ string theory. At tree level, open $N{=}2$ strings are known to 
reduce to the ASDYM model in a Lorentz noninvariant gauge~\cite{n2old}; 
their SFT formulation~\cite{Be95} is based on the $N{=}4$ topological string 
description~\cite{BeVa94,OoVa95}. The latter contains twistors from the outset:
The coordinate $\z\in\C P^1$, the linear system, the integrability and the 
classical solutions with the help of twistor methods were all incorporated 
into $N{=}2$ open string field theory in~\cite{LePo00,LePoUh}.
Since this theory~\cite{Be95} generalizes the Wess-Zumino-Witten-type model
\cite{NaSch92} for ASDYM theory and thus describes only anti-self-dual gauge
fields (having helicity $+1$), it is not Lorentz invariant.
Its maximally supersymmetric extension, \nfour\ super ASDYM theory, however,
does admit a Lorentz-invariant formulation~\cite{Si92,ChaSi96}. This theory
and its truncation to \nzero\ features pairs of fields of opposite helicity. 
In~\cite{BeSi97} it was proposed to lift the corresponding Lagrangians to SFT.

In the present paper the twistor description of both the purely bosonic 
and the \nfour\ supersymmetric ASDYM models~\cite{Si92,ChaSi96} is raised to 
the SFT level. In contrast to previous proposals~\cite{Witten}--\cite{BeWi},
we allow the string to vibrate only in part of the supertwistor space.
The remaining coordinates of this space are not promoted to word-sheet fields
but kept as non-dynamical string field parameters. Concretely,
we propose a cubic action containing an integration over the supertwistor 
space $\C P^{3|4}$ and show that its hCS-type equations of motion are 
equivalent to the covariant string field equations introduced in \cite{BeSi97}.
This model may be regarded as a specialization of Witten's supertwistor SFT
and may even be equivalent to it.
In any case, it is directly related with \nfour\ super ASDYM theory in four 
dimensions. We also consider its proper truncation to the bosonic sector, 
which yields a twistor SFT related to non-supersymmetric ASDYM theory.

\section{Covariant string field theory for open N=2 strings}

\noindent
{\bf Open N=2 strings.\ }
{}From the worldsheet point of view critical open $N{=}2$ strings in a flat 
four-dimensional space-time of signature $({-}\,{-}\,{+}\,{+})$ or 
$({+}\,{+}\,{+}\,{+})$ are nothing but $N{=}2$ supergravity 
on a two-dimensional (pseudo) Riemannian surface with 
boundaries, coupled to two chiral $N{=}2$ matter multiplets $(X,\psi)$. 
The latter's components are complex scalars (the four imbedding coordinates)
and Dirac spinors (their four NSR partners) in two dimensions. 
In the ghost-free formulation of the $N{=}2$ string one employs the extension 
of the $c{=}6, N{=}2$ superconformal algebra to the ``small'' $N{=}4$ 
superconformal algebra\footnote{
We raise and lower indices with 
$\eps^{12}=-\eps^{21}=-1$, $\eps_{12}=-\eps_{21}=1$, 
and similarly for $\eps^{\ad\bd}, \eps_{\ad\bd}$ and 
$\eps^{\add\bdd}, \eps_{\add\bdd}$.} 
\beq \bal
T&\=\pa_z X^{\a\bd}\,\pa_z X_{\a\bd}\ +\ \psi^{\ad\bdd}\,\pa_z\psi_{\ad\bdd}\ ,
\\[4pt]
G_\a^{\ \ \bdd} &\= \psi^{\gd\bdd}\,\pa_z X_{\a\gd}\ ,\qquad
J^{\add\bdd} \= \psi^{\gd\add}\,\psi_\gd^{\ \ \bdd}\ ,
\eal \eeq
where $\a,\b=1,2$ and $\ad,\bd=\od,\zd$ are space-time spinor indices and
$\add,\bdd=\odd,\zdd$ denote the world-sheet internal indices associated
with the group SU(1,1)${}''$ (Kleinian space $\R^{2,2}$) or SU(2)${}''$ 
(Euclidean space $\R^{4,0}$) of R~symmetries. 
For the reality structures imposed on target space coordinates and 
superconformal algebra generators see~\cite{BeVa94,Be95,BeVaWi99,LePoUh}.  
After twisting this algebra,
\beq
\Db_\a \ :=\ G_\a^{\ \ \odd}
\eeq
become two fermionic spin-one operators which subsequently serve as 
BRST-like currents since they are nilpotent~\cite{BeVa94,Be95},
\beq \label{nilpotent}
(\Db_1)^2 \= 0 \= (\Db_2)^2  \und  \bigl\{\Db_1\,,\,\Db_2\bigr\} \= 0\ .
\eeq
Furthermore, $\psi^{\ad\odd}$ is now conformal spin zero while
$\psi^{\ad\zdd}$ is conformal spin one.

\bigskip

\noindent
{\bf Covariant string field theory.\ }
Following Berkovits and Siegel~\cite{BeSi97}, we introduce two Lie-algebra 
valued fermionic string fields $A_\a[X,\psi]$ and three Lie-algebra valued 
bosonic string fields $G^{\a\b}[X,\psi]$ (symmetric in $\a$ and~$\b$).
Although we suppress it in our notation, string fields are always multiplied
using Witten's star product (midpoint gluing prescription) 
\cite{Wi85}.\footnote{ 
The star product was concretized in oscillator language for bosons in
\cite{GrJe87} and for twisted fermions in \cite{KlUhl}.}
The index structure reveals that the fields $G_{\a\b}$ parametrize the 
self-dual\footnote{
Self-duality can always be interchanged with anti-self-duality by flipping
the orientation of the four-dimensional target space. For the choice of the
orientation made in~\cite{Witten,PoSae} these $G_{\a\b}$ parametrize a 
self-dual tensor.}
tensor $G_{\a\ad,\b\bd}=\eps_{\ad\bd}G_{\a\b}$ on the target space.

The Lorentz invariant string field theory action~\cite{BeSi97} reads
\beq \label{SBS1}
S_{\text{BS}} \= \<\,\tr\,(G^{\a\b}\,F_{\a\b})\,\>\ ,
\eeq
where $\<\dots\>$ means integration over all modes of $X$ and~$\psi$, 
the trace ``$\tr$'' is taken over the Lie algebra indices and
\beq
F_{\a\b}\ :=\ \Db_\a A_\b\ +\ \Db_\b A_\a\ +\ \bigl\{ A_\a\,,A_\b\bigr\}\ .
\eeq
Note that the action of $\Db_\a$ on any string field~$B$ is defined
in conformal field theory language as taking the contour integral~\cite{Be95}
\beq \label{contour}
\bigl(\Db_\a B\bigr)(z)\=\oint_z\,\frac{\diff w}{2\pi\im}\ \Db_\a(w)\,B(z)\ .
\eeq
The covariant string field equations of motion following from the action
(\ref{SBS1}) read
\beq \label{N0eom}
F_{\a\b}\=0  \und  \Db_\a G^{\a\b}\ +\ \bigl[ A_\a\,,G^{\a\b} \bigr] \= 0\ .
\eeq

For a supersymmetric generalization of the action~(\ref{SBS1}) 
Berkovits and Siegel~\cite{BeSi97} introduce a multiplet of string fields
\beq
\bigl( A_\a\,,\,\ct_i\,,\,\phi_{ij}\,,\,\c^{\a i}\,,\,G^{\a\b} \bigr)
\qquad\text{with}\quad i,j=1,2,3,4
\eeq
imitating the \nfour\ ASDYM multiplet~\cite{Si92}.
Here, $A_\a$ and $\c^{\a i}$ are fermionic while $\ct_i$, $\phi_{ij}$ and
$G^{\a\b}$ are bosonic. Ref.~\cite{BeSi97} proposes the following action for
this super SFT:
\beq \label{SBS2}
\widehat{S}_{\text{BS}} \= \<\,\tr\,\bigl( 
G^{\a\b}\,F_{\a\b}\ +\ 2\,\c^{\a i}\,\nabla_\a\ct_i\ +\ 
\sfrac18 \eps^{ijkl} \phi_{ij}\,\nabla_\a\nabla^\a \phi_{kl}\ +\
\sfrac12 \eps^{ijkl} \phi_{ij}\, \ct_k\, \ct_l \bigr)\,\>
\eeq
with
\beq
\nabla_\a B \ :=\ \Db_\a B\ +\ A_\a B\ -\ ({-}1)^{|B|} B A_\a\ ,
\eeq
where the Grassmann parity $|B|$ equals 0 or 1 for bosonic or fermionic 
fields~$B$, respectively.
Due to the large number of string fields this model seems unattractive.
However, as we shall see in the coming section, all these fields appear as
components of {\em one\/} string field living in a twistor extended target 
space.

\section{Cubic string field theory for open N=2 strings}

\noindent
{\bf Supertwistor space notation.\ } 
In the Appendix we describe the supertwistor space $\cP_\eps^{3|4}$
of the space $(\R^4,g_\eps)$ with the metric
$g_\eps=\text{diag}(-\eps,-\eps,1,1)$ and $\eps=\pm1$. It is fibered over
the real two-dimensional space $\Sigma_\eps$ with $\Sigma_{-1}=\C P^1$ and
$\Sigma_{+1}=H^2$ covered by two patches $U^\eps_\pm$. 
The space $\cP_\eps^{3|4}$ is parametrized by four even complex coordinates
$(x^{\a\ad})\in\C^4$ subject to the reality conditions
$x^{2\zd}=\bar{x}^{1\od}$ and $x^{2\od}=\eps\bar{x}^{1\zd}$,
complex coordinates $\z_\pm\in U^\eps_\pm$ and odd (Grassmann) coordinates
$\t^i_\pm$, $i=1,\dots,4$. 
The space $\cP_\eps^{3|4}$ is a Calabi-Yau supermanifold~\cite{Witten}. 
{}From now on we shall work on the patch $U^\eps_+$ of $\Sigma_\eps$, 
and for notational simplicity we shall omit the subscript ``+'' in 
$\z_+\in U^\eps_+$, $\t^i_+$ etc. For further use we introduce
\beq \label{zetadef}
(\z_\a)= \Bigl(\begin{matrix}  1 \\ \z \end{matrix}\Bigr)\ ,\qquad
(\z^\a)= \Bigl(\begin{matrix} -\z \\ 1 \end{matrix}\Bigr)\ ,\qquad
(\zh^\a)=\Bigl(\begin{matrix}\eps\\-\zb\end{matrix}\Bigr) \und
\n = (1-\eps\,\z\zb)^{-1}\ .
\eeq

\bigskip

\noindent
{\bf BRST operator.\ }
Let us introduce the operator
\beq \label{Dbar}
\Db\ :=\ \z^\a\,\Db_\a \= \psi^{\bd\odd}\,\pa_z (\z^\a\,X_{\a\bd})
\eeq
taking values in the holomorphic line bundle $\cO(1)$. We notice that
the operators $\z^\a\pa_z X_{\a\bd}$ act as derivatives on string fields.
Their zero mode parts, $\z^\a\frac{\pa}{\pa x^{\a\bd}}$, form two
type (0,1) vector fields on the bosonic twistor space $\cP_\eps^3$ 
fibered over $\Sigma_\eps$ (see the Appendix for more details).
Recall that $\cP_\eps^3$ being an open subset of $\C P^3$ is the
twistor space of $(\R^4,g_\eps)$. In order to obtain a general
type (0,1) vector field along the twistor space one should therefore
extend the operator~(\ref{Dbar}) by adding the type (0,1) derivative
along~$\Sigma_\eps$, 
\beq
\pab\ :=\ \diff\zb\,\frac{\pa}{\pa\zb} \ .  
\eeq
Assuming that string fields now depend on the extra variable 
$\z\in\Sigma_\eps$, we define the operator
\beq \label{BRST}
Q\ :=\ \Db\ +\ \pab
\eeq
acting on string fields via (\ref{contour}) for the $\Db$ part and by ordinary
differentiation with respect to~$\zb$. It is easy to see that $Q^2=0$
due to (\ref{nilpotent}) and the facts that $\Db$ does not depend on~$\zb$ 
and that $\{\diff\zb,\psi^{\ad\bdd}\}=0$ (cf.~\cite{Berezin}). We take this 
nilpotent operator as the BRST operator of our SFT extended to
$\Sigma_\eps^{1|4}\hookrightarrow\cP_\eps^{3|4}$.

\bigskip

\noindent
{\bf String fields.\ }
We now consider a fermionic (odd) string field $\cA[X,\psi,\t^i,\z,\zb]$
depending not only on $X(\s)$ and $\psi(\s)$ but also on $\t^i$ and on the
parameter $\z\in\Sigma_\eps$. It is important to realize that $\t^i$ and $\z$
do not depend on~$\s$ here but may be considered as zero modes of world-sheet
fields. Since the operator $Q$ has the split form (\ref{BRST}) it is natural 
to assume the same splitting of the string field~$\cA$,
\beq \label{Asplit}
\cA \= \cA_{\Db}[X,\psi,\t^i,\z,\zb]\ +\ \cA_{\pab}[X,\psi,\t^i,\z,\zb]
\qquad\text{with}\qquad \cA_{\pab} = \cA_{\zb}\,\diff\zb \ ,
\eeq
where $\cA_{\Db}$ gauges $\Db$ and $\cA_{\pab}$ gauges $\pab$. 
Note also that $\Db$ takes values in $\cO(1)$ and $\pab$ in $\cO(0)$; 
therefore, $\cA_{\Db}$ and $\cA_{\pab}$ are $\cO(1)$ and $\cO(0)$ valued, 
respectively. Since $\diff\zb$ is the basis section of the bundle 
$\bar{\cO}(-2)$ complex conjugate to $\cO(2)$
and anticommutes with spinors $\psi^{\ad\bdd}$ we reason that $\cA_{\zb}$ is 
bosonic (even) and takes values in $\bar{\cO}(-2)$. It is also assumed that a 
term $\cA_\z\diff\z$ is absent in the splitting (\ref{Asplit}), i.e.~$\cA$ is a 
string field of the (0,1)-form type (cf.~\cite{Witten} for the B-model 
argument). Note that by definition the string field $\cA_{\Db}$ does not 
contain $\diff\zb$ and is fermionic (odd).

\bigskip

\noindent
{\bf Cubic action.\ }
Having $\diff\z$ and $\diff\t^i$ we introduce the action
\beq \label{N4S}
S \= \int\!\diff\z
\int\!\diff\t^1\,\diff\t^2\,\diff\t^3\,\diff\t^4\
\<\,\tr\,\bigl( \cA\,Q\cA \ +\ \sfrac23 \cA^3 \bigr) \,\> \ ,
\eeq
where $\<\dots\>$ is the same integration over $(X,\psi$) modes as in
(\ref{SBS1}). Note that $\int\!\diff\z$ acts as integration over $\Sigma_\eps$
for terms containing $\diff\zb$ and as a contour integral around $\z{=}0$
for other terms. 
The Lagrangian~$\cL$ in~(\ref{N4S}) can be split into two parts,
\bea \label{lag}
\cL &\=& \tr\,\bigl( \cA\,Q\cA \ +\ \sfrac23 \cA^3 \bigr) \= \cL_1 + \cL_2 \ ,
\qquad\text{with} \\[10pt] \label{lag1}
\cL_1 &\=& \tr\,\bigl( \cA_\Db\,\pab\cA_\Db \ +\ 2\,\cA_\Db\,\Db\cA_\pab \ +\
          2\,\cA_\pab\,\cA_\Db^2 \bigr) \ , \\[4pt] \label{lag2}
\cL_2 &\=& \tr\,\bigl( \cA_\Db\,\Db\cA_\Db \ +\ \sfrac23 \cA_\Db^3 \bigr)\ ,
\eea
where we used the cyclicity under the trace and omitted total derivatives.

It is important to note that $\cL_1$ takes values in $\cO(2)$, which is 
compensated by the holomorphic measure 
$\diff\z\,\diff\t^1\,\diff\t^2\,\diff\t^3\,\diff\t^4$ 
being $\cO(-2)$ valued.\footnote{
The choice of four Grassmann coordinates $\t^i$ is dictated by the 
Calabi-Yau condition: The contribution of the coordinates $(X,\z,\t)$ 
to the first Chern number is $(2,2,-4)$, respectively.}
At the same time, $\cL_2$ takes values in $\cO(3)$ which causes it to drop out
of the action by virtue of Cauchy's theorem applied to the $\z$ contour 
integral. Thus, the action (\ref{N4S}) can be rewritten as
\beq \label{N4S2}
S \= \int\!\diff\z
\int\!\diff\t^1\,\diff\t^2\,\diff\t^3\,\diff\t^4\
\<\,\tr\,\bigl( \cA_\Db\,\pab\cA_\Db \ +\ 2\,\cA_\Db\,\Db\cA_\pab \ +\
2\,\cA_\pab\,\cA_\Db^2 \bigr)\,\> \ .
\eeq
Moreover, both forms (\ref{N4S}) and (\ref{N4S2}) of the action lead to the 
same Chern-Simons-type equation of motion,
\beq \label{eom}
Q\,\cA \ +\ \cA^2 \= 0
\eeq
which decomposes into
\bea \label{eom1}
\Db\cA_\pab\ +\ \pab\cA_\Db\ +\ \bigl\{\cA_\Db\,,\cA_\pab\bigr\}&\=&0\\[4pt]
\text{and}\qquad\qquad \Db\cA_\Db\ +\ \cA_\Db^2 &\=& 0 \ . \label{eom2}
\eea

\bigskip

\noindent
{\bf Component analysis.\ }
Recall that $\cA_\Db$ and $\cA_\zb$ take values in the bundles $\cO(1)$ and
$\bar{\cO}(-2)$, respectively. Together with the fact that the $\t^i$ are 
nilpotent and $\cO(1)$ valued, this determines the dependence of $\cA_\Db$ and
$\cA_\pab=\cA_\zb\diff\zb$ on $\t^i$, $\z$ and $\zb$. 
Namely, this dependence has the form (cf.~\cite{PoSae})
\beq \label{N4exp} \bal
\cA_\Db &\= \z^\a A_\a + \t^i\ct_i\ +\ 
\sfrac{\n}{2!}\t^{ij}\zh^\a\p_{\a ij}\ +\
\sfrac{\n^2}{3!}\t^{ijk}\zh^\a\zh^\b\c_{\a\b ijk}\ +\
\sfrac{\n^3}{4!}\t^{ijkl}\zh^\a\zh^\b\zh^\g G_{\a\b\g ijkl} \ ,\\[8pt]
\cA_\zb &\= \sfrac{\n^2}{2!}\t^{ij}\p_{ij}\ +\
\sfrac{\n^3}{3!}\t^{ijk}\zh^\a\c_{\a ijk}\ +\
\sfrac{\n^4}{4!}\t^{ijkl}\zh^\a\zh^\b G_{\a\b ijkl} \ ,
\eal \eeq
where $\z^\a$, $\zh^\a$ and $\nu$ are given in (\ref{zetadef})
and $\ \t^{i_1i_2\dots i_k}:=\t^{i_1}\t^{i_2}\dots\t^{i_k}$. The expansion 
(\ref{N4exp}) is defined up to a gauge transformation generated by
a group-valued function which may depend on $\z$ and~$\zb$. All string fields
appearing in the expansion (\ref{N4exp}) depend only on $X(\s)$ and~$\psi(\s)$.
{}From the properties of $\cA_\Db$, $\cA_\pab$ and $\t^i$ it follows that
the fields with an odd number of spinor indices are fermionic (odd) while
those with an even number of spinor indices are bosonic (even).
Moreover, due to the symmetry of the $\zh^\a$ products and the skewsymmetry
of the $\t^i$ products all component fields are automatically symmetric in 
their spinor indices and antisymmetric in their Latin indices.

Substituting (\ref{N4exp}) into (\ref{eom1}), we obtain the equations\footnote{
Round brackets denote symmetrization with respect to enclosed indices.}    
\beq
\p_{\a ij} \= -\nabla_\a \p_{ij} \und
\c_{\a\b ijk} \= \sfrac12 \nabla_{\!(\a} \c_{\b)ijk} \und
G_{\a\b\g ijkl} \= -\sfrac13 \nabla_{\!(\a} G_{\b)\g ijkl} 
\eeq
showing that $(\p_{\a ij}, \c_{\a\b ijk}, G_{\a\b\g ijkl})$ 
is a set of auxiliary fields.
The other nontrivial equations following from (\ref{eom1}) and (\ref{eom2}) 
after substituting (\ref{N4exp}) read\footnote{
Recall that $\eps=1$ for signature $({-}\,{-}\,{+}\,{+})$ and $\eps=-1$ for 
signature $({+}\,{+}\,{+}\,{+})$.}    
\beq \label{eomcomp} \bal
F_{\a\b}\ \equiv\ 
\Db_\a A_\b\ +\ \Db_\b A_\a\ +\ \bigl\{A_\a\,,A_\b\bigr\}&\=0\ ,\\[4pt]
\nabla_\a \ct_i &\= 0 \ ,\\[4pt]
\eps^{\a\b}\nabla_\a\c^i_\b\ +\ 2\eps\,\bigl[\p^{ij}\,,\ct_j\bigr]&\=0\ ,\\[4pt]
\nabla_\a \nabla^\a \p_{ij}\ +\ 2\eps\,\bigl[\ct_i\,,\ct_j\bigr] &\=0\ ,\\[4pt]
\eps^{\a\b}\nabla_\a G_{\b\g}\ +\ 2\eps\,\bigl[\ct_i\,,\c^i_\g\bigr]\ +\
  \eps\,\bigl[\nabla_\g\p_{ij}\,,\p^{ij}\bigr] &\=0\ ,
\eal \eeq  
where we introduced 
\beq \label{dual}
\p^{ij}\ :=\ \sfrac{1}{2!}\eps^{ijkl}\p_{kl} \und
\c^i_\a\ :=\ \sfrac{1}{3!}\eps^{ijkl}\c_{\a jkl} \und
G_{\a\b}\ :=\ \sfrac{1}{4!}\eps^{ijkl}G_{\a\b ijkl}\ .
\eeq

Up to constant field rescalings 
\beq \label{rescale}
G_{\a\b}\ \to\ -G_{\a\b}\ ,\qquad
\ct_i\ \to\ \sfrac12\ct_i\ ,\qquad
\p_{ij}\ \to\ \sfrac12\p_{ij} \und
\c^i_\a\ \to\ \c^i_\a
\eeq
the equations (\ref{eomcomp}) for $\eps{=}1$
coincide with the equations of motion following from the action (\ref{SBS2})
proposed by Berkovits and Siegel~\cite{BeSi97}. In the zero mode sector they
reduce to the anti-self-dual \nfour\ super Yang-Mills equations of motion.
Hence, we have established that maximally supersymmetric ASDYM theory can
be obtained from the standard cubic SFT for a single string field~$\cA$
after extending the setting to the supertwistor space.

\section{Bosonic truncation of open string field theory}

In order to make contact with non-supersymmetric ASDYM theory,
we subject our string field $\cA$ from (\ref{Asplit}) and (\ref{N4exp}) 
to the truncation conditions
\beq \label{trunc}
\int\!\diff\t^1\,\diff\t^2\,\diff\t^3\,\diff\t^4\ \Biggl\{
\begin{matrix} \t^i \\ \t^{ij} \\ \t^{ijk} \end{matrix} \Biggr\}\,\cA \=0 \ .
\eeq
These conditions imply that $\cA$ depends only on the combination
\beq
\t\ :=\ \t^1\t^2\t^3\t^4\ ,\qquad\text{i.e.}\quad
\cA \= \cA[X,\psi,\t,\z,\zb]\ .
\eeq
Obviously, the even nilpotent variable~$\t$ belongs to the bundle $\cO(4)$ and 
the integration measure in (\ref{trunc}) to $\cO(-4)$.

The properties of the truncated string field
\beq
\cA \= \cA_\Db[X,\psi,\t,\z,\zb] \ +\ \cA_\pab[X,\psi,\t,\z,\zb]
\eeq	
are the same as before the truncation, except for the restricted dependence
on the Grassmann variables. The operators $Q$, $\Db$ and $\pab$, the
actions (\ref{N4S}) and (\ref{N4S2}), the Lagrangians (\ref{lag})--(\ref{lag2})
and the equations of motion (\ref{eom})--(\ref{eom2}) are unchanged. However, 
the expansion (\ref{N4exp}) now simplifies to
\beq \label{N0exp} \bal
\cA_\Db &\=  \z^\a\,A_\a[X,\psi] \ +\ 
             \t\,\n^3\zh^\a\zh^\b\zh^\g\,G_{\a\b\g}[X,\psi] \ ,\\[4pt]
\cA_\pab &\= \t\,\n^4\zh^\a\zh^\b\,G_{\a\b}[X,\psi]\,\diff\zb \ ,
\eal \eeq
where (see (\ref{dual}))
\beq
G_{\a\b}\ =\ \sfrac{1}{4!}\eps^{ijkl}G_{\a\b ijkl} \und
G_{\a\b\g}\ :=\ \sfrac{1}{4!}\eps^{ijkl}G_{\a\b\g ijkl} \ .
\eeq
{}From the properties of $\cA_\Db$ and $\cA_\pab$ it follows that the
string fields $A_\a$ and $G_{\a\b\g}$ are odd and the $G_{\a\b}$ are even.

Substituting the expansion (\ref{N0exp}) into the equations of motion 
(\ref{eom1}) and (\ref{eom2}), we recover for $A_\a$ and $G_{\a\b}$ the bosonic
string field equations 
\beq \label{N0eom2}
F_{\a\b}\=0  \und  \Db_\a G^{\a\b}\ +\ \bigl[ A_\a\,,G^{\a\b} \bigr] \= 0
\eeq
displayed already in (\ref{N0eom}) 
and for $G_{\a\b\g}$ the dependence
\beq
G_{\a\b\g} \= -\sfrac{1}{3}\,\nabla_{(\a} G_{\b)\g} 
\eeq
as expected. The same result occurs when putting to zero in (\ref{eomcomp})
the string fields $\ct_i$, $\c^i_\a$ and $\p_{ij}$ as the truncation 
(\ref{trunc}) demands.
All other equations following from (\ref{eom1}) and (\ref{eom2}) are satisfied 
automatically, due to (\ref{N0eom2}) and the Bianchi identities. 

Hence, we have proven that the cubic supertwistor SFT defined by the action
(\ref{N4S}) together with the geometric truncation conditions~(\ref{trunc}) is 
equivalent to the Berkovits-Siegel SFT given by the action~(\ref{SBS1}).
Moreover, (\ref{SBS1}) and (\ref{SBS2}) derive from (\ref{N4S}) simply by 
substituting there the expansion (\ref{N0exp}) or (\ref{N4exp}), respectively,
and integrating over the Grassmann and twistor variables.
All this is similar to the field theory case~\cite{Witten,PoSae} where in the
supertwistor reformulation of \nfour\ ASDYM theory as hCS theory
the dependence of all fields on the twistor variable~$\z$ is fixed 
(up to a gauge transformation) by the topology of the supertwistor space
and one can integrate over it, descending from six to four real dimensions.

\section{Conclusions}

The basic result of this paper can be summarized in the equations
\bea
S &=& \int\!\diff\z 
\int\!\diff\t^1\,\diff\t^2\,\diff\t^3\,\diff\t^4\
\<\,\tr\,\bigl( \cA\,Q\cA \ +\ \sfrac23 \cA^3 \bigr) \,\> \\[4pt]
&=& \int_{\Sigma_\eps}\!\!\!\diff\z{\wedge}\diff\zb
\int\!\diff\t^1\,\diff\t^2\,\diff\t^3\,\diff\t^4\
\<\,\tr\,\bigl( -\cA_\Db\,\pa_\zb\cA_\Db \ +\ 2\,\cA_\Db\,\Db\cA_\zb \ +\
2\,\cA_\Db^2\,\cA_\zb \bigr)\,\> \\[8pt]
&=& \ c\ \<\,\tr\,\bigl(
G^{\a\b}\,F_{\a\b}\ +\ 2\eps\,\c^{\a i}\,\nabla_\a\ct_i\ +\
\sfrac14\eps\,\phi^{ij}\,\nabla_\a\nabla^\a \phi_{ij}\ +\ 
\phi^{ij}\, \ct_k\, \ct_l \bigr)\,\> \ ,
\eea
where $c$ is an inessential numerical constant.
The first step demands a split $\cA=\cA_\Db+\cA_\zb\diff\zb$ of the basic 
(supertwistor) string field. The second step requires integrating over 
$\Sigma_\eps^{1|4}$ and rescaling the field as in~(\ref{rescale}).
Truncating $\cA$ to its lowest and highest Grassmann components
(the $O(\t^0)$ and $O(\t^4)$ parts) projects the above action to 
$\<\tr(G^{\a\b}F_{\a\b})\>$, which governs bosonic $N{=}2$ open SFT.
Finally, reducing to the string zero modes one recovers the twistor description
of \nfour\ and \nzero\ ASDYM on the field theory level.

\bigskip

\noindent
{\bf Acknowledgements}

This work was partially supported by the Deutsche Forschungsgemeinschaft (DFG).

\begin{appendix} 
\section{Appendix: Supertwistor space}

\noindent
{\bf The twistor space of $\R^{4,0}$.\ }
Let us consider the Riemann sphere $\C P^1\cong S^2$ with homogeneous 
coordinates $(\mu_\a)\in\C^2$. It can be covered by two patches,
\beq
U_+ \= \bigl\{ (\mu_1,\mu_2):\ \mu_1\not=0 \bigr\} \und
U_- \= \bigl\{ (\mu_1,\mu_2):\ \mu_2\not=0 \bigr\}
\eeq
with coordinates $\z_+:=\mu_2/\mu_1$ on $U_+$ and $\z_-:=\mu_1/\mu_2$ on $U_-$.
On the intersection $U_+\cap U_-$ we have $\z_+=\z_-^{-1}$.

The holomorphic line bundle $\cO(n)$ over $\C P^1$ is defined as a 
two-dimensional complex manifold with the holomorphic projection
\beq
\pi:\quad \cO(n)\ \to\ \C P^1
\eeq
such that it is covered by two patches $\tilde{U}_+$ and $\tilde{U}_-$ with
coordinates $(w_+,\z_+)$ on $\tilde{U}_+$ and $(w_-,\z_-)$ on $\tilde{U}_-$
related by $w_+=\z_+^nw_-$ and $\z_+=\z_-^{-1}$ on 
$\tilde{U}_+\cap\tilde{U}_-$.
A global holomorphic section of $\cO(n)$ exists only for $n{\geq}0$.
Over $U_\pm\subset\C P^1$ it is represented by polynomials $p_\pm^{(n)}$ in
$\z_\pm$ of degree~$n$ with $p_+^{(n)}=\z_+^n p_-^{(n)}$ on $U_+\cap U_-$.

Recall that the Riemann sphere 
\beq
\C P^1 \ \cong\ \text{SO(4)}/\text{U(2)}
\eeq
parametrizes the space of all translational invariant (constant) complex 
structures on the Euclidean space $\R^{4,0}$, and the space
$\cP_E^3:=\R^4{\times}\C P^1$ is called the twistor space of~$\R^{4,0}$
\cite{AHS}. As a complex manifold $\cP_E^3$ is a rank~2 holomorphic vector 
bundle $\cO(1)\oplus\cO(1)$ over $\C P^1$:
\beq \label{P3E}
\cP_E^3 \= \cO(1)\oplus\cO(1) \ .
\eeq
For more details and references see e.g.~\cite{AHS,Po98,PoSae}.

\bigskip

\noindent
{\bf The twistor space of $\R^{2,2}$.\ }
In the Kleinian space $\R^{2,2}$ of signature $({-}\,{-}\,{+}\,{+})$ 
constant complex structures are parametrized by the two-sheeted hyperboloid
\beq
H^2 \= H_+ \cup H_- \ \cong\ \text{SO}(2,2)/\text{U}(1,1) \ ,
\eeq
where 
\beq \bal
H_+&\=\bigl\{\z_+\in U_+:\ |\z_+|<1\bigr\}\ \cong\ \text{SU}(1,1)/\text{U}(1)\\
\und 
H_-&\=\bigl\{\z_-\in U_-:\ |\z_-|<1\bigr\}\ \cong\ \text{SU}(1,1)/\text{U}(1)
\ .
\eal \eeq
In fact, under the action of the group SU(1,1) the Riemann sphere is decomposed
into three orbits, $\C P^1=H_+\cup S^1\cup H_-$, where the boundary of both
$H_+$ and $H_-$ is given by
\beq
S^1\=\bigl\{\z\in\C P^1:\ |\z|=1\bigr\}\ \cong\ \text{SU}(1,1)/B_+ 
\qquad\text{with}\quad B_+\=\Bigl(\begin{smallmatrix}
a_1+\im a_2 \ {}&{}\ a_3-\im a_2 \\[6pt] a_3+\im a_2 \ {}&{}\ a_1-\im a_2 
\end{smallmatrix}\Bigr)
\eeq
with $a_{1,2,3}\in\R$ and $a_1^2{-}a_2^2=1$.

The twistor space of $\R^{2,2}$ is the space $\cP_K^3:=\R^4{\times}H^2$ which
as a complex manifold coincides with the restriction of the rank~2 holomorphic
vector bundle~(\ref{P3E}) to the bundle over $H^2\subset\C P^1$. Equivalently,
it can be described as a space $\cP_K^3=\cP_E^3\setminus\cT^3$, where $\cT^3$
is a real three-dimensional subspace of $\cP_E^3$ stable under an anti-linear
involution (real structure) which can be defined on $\cP_E^3$. For more details
see e.g.~\cite{n2hidden}.

\bigskip

\noindent
{\bf Vector fields of type (0,1).\ }
For considering both signatures together, we denote by $\Sigma_\eps$ the space 
of complex structures on $(\R^4,g_\eps)$ with the metric 
$g_\eps=\text{diag}({-}\eps,{-}\eps,1,1)$ and $\eps{=}\pm1$, so that
\beq
\Sigma_{-1} \= \C P^1 \und \Sigma_{+1} \= H^2 \ .
\eeq
Therefore, $\Sigma_\eps$ is covered by two patches $U_\pm^\eps$ with
$U_\pm^{-1}=U_\pm$ and $U_\pm^{+1}=H_\pm$. Analogously, we denote by
$\cP^3_\eps$ the twistor space of $(\R^4,g_\eps)$ with $\cP^3_{-1}=\cP^3_E$
and $\cP^3_{+1}=\cP^3_K$. The complex manifold $\cP^3_\eps$ is covered by two
patches $\cal{V}_\pm^\eps$ with complex coordinates $(w_\pm^\ad,\z_\pm)$ on
$\cal{V}_\pm^\eps$. We introduce
\beq
(\z^\pm_\a)= \Bigl(\begin{matrix}  1 \\ \z_\pm \end{matrix}\Bigr)\ ,\quad
(\z^\a_\pm)= \Bigl(\begin{matrix} -\z_\pm \\ 1 \end{matrix}\Bigr)\ ,\quad
\n_+ = (1-\eps\,\z_+\zb_+)^{-1} \quad\text{and}\quad 
\n_- = -\eps\,(1-\eps\,\z_-\zb_-)^{-1}
\eeq
for $\z_\pm\in U_\pm^\eps$. 
Note that in terms of $(\z^\pm_\a)$ or $(\z^\a_\pm)$
a section of the bundle $\cO(n)$ over $U_\pm^\eps$ can be written as 
\beq
p_\pm^{(n)} \= p^{\a_1\dots\a_n}\,\z^\pm_{\a_1}\dots\z^\pm_{\a_n}
\= p_{\a_1\dots\a_n}\,\z_\pm^{\a_1}\dots\z_\pm^{\a_n} \ .
\eeq

Recall that $(\R^4,g_\eps)$ can be parametrized by coordinates 
$(x^{\a\ad})\in\C^4$ with the reality conditions $x^{2\zd}=\bar{x}^{1\od}$ and
$x^{2\od}=\eps\,\bar{x}^{1\zd}$~\cite{PoSae}. On the twistor space 
$\cP^3_\eps\cong\R^4{\times}\Sigma_\eps$ we have coordinates 
$(w_\pm^\ad,w_\pm^{\dot{3}})=(x^{\a\ad}\z^\pm_\a,\z_\pm)$ or
$(x^{\a\ad},\z_\pm,\zb_\pm)$. The antiholomorphic vector fields
$\pa/\pa\bar{w}_\pm^\ad$ and $\pa/\pa\bar{w}_\pm^{\dot{3}}$ can be rewritten
in the coordinates $(x^{\a\ad},\z_\pm,\zb_\pm)$ as
\beq
\frac{\pa}{\pa\bar{w}_\pm^\od}=\n_\pm\z_\pm^\a\frac{\pa}{\pa x^{\a\zd}}\ ,
\qquad
\frac{\pa}{\pa\bar{w}_\pm^\zd}=\eps\,\n_\pm\z_\pm^\a\frac{\pa}{\pa x^{\a\od}}
\und
\frac{\pa}{\pa\bar{w}_\pm^{\dot{3}}}=
\frac{\pa}{\pa\zb_\pm}-\eps\,x^{1\ad}\z^\a\frac{\pa}{\pa x^{\a\ad}} \ .
\eeq
The vector fields
\beq
\bar{v}^\pm_\ad \= \z^\a_\pm\,\frac{\pa}{\pa x^{\a\ad}} \und
\bar{v}^\pm_{\dot{3}} \= \frac{\pa}{\pa\zb_\pm}
\eeq
can be taken as a basis of vector fields of type (0,1) on~$\cP^3_\eps$.

\bigskip 

\noindent
{\bf Supertwistors.\ }
Let us now add four odd variables $\t^i$ such that
\beq
\t^i\,\t^j \ +\ \t^j\,\t^i \= 0 \qquad\text{for}\quad i,j=1,2,3,4
\eeq
and each $\t^i$ takes its value in the line bundle $\cO(1)$ over $\C P^1$
(cf.~\cite{Witten}). For describing them formally we introduce a Grassmann
parity changing operator~$\Pi$ which, when acting on a vector bundle, flips
the Grassmann parity of the fibre coordinates. Hence, we consider the bundle
$\Pi\,\C^4\otimes\cO(1)\to\C P^1$
which is parametrized by complex variables $\z_\pm\subset U_\pm\subset\C P^1$
and fibre Grassmann coordinates~$\t^i_\pm$ such that $\t^i_+=\z_+\t^i_-$ on
the intersection of the two patches covering the total space of this vector
bundle.

With the Grassmann variables~$\t^i$ one can introduce the supertwistor space
$\cP_E^{3|4}$ as a holomorphic vector bundle over~$\C P^1$, namely
\beq
\cP_E^{3|4} \= \C^2\otimes\cO(1)\ \oplus\ \Pi\,\C^4\otimes\cO(1)\ .
\eeq
The supertwistor space $\cP_K^{3|4}$ is defined as a restriction
of the bundle $\cP_E^{3|4}\to\C P^1$ to the bundle over the
two-sheeted hyperboloid $H^2\subset\C P^1$.

\end{appendix}

\end{document}